\documentstyle[11pt,paspconf]{article}
\input epsf.tex
\begin{document}

\title{Weighing Superclusters}

\author{Chung--Pei Ma}
\affil{Department of Physics and Astronomy, University of Pennsylvania.
    Philadelphia, PA 19104}

\author{Todd Small}
\affil{Institute of Astronomy,
University of Cambridge, Madingley Road, Cambridge CB3 0HA, UK}

\author{Wallace Sargent}
\affil{Division of Physics, Mathematics, and Astronomy, California 
Institute of Technology, Pasadena, CA 91125}

\begin{abstract}
We present a summary report on a detailed study of the mass and
dynamical state of the Corona Borealis supercluster using the largest
redshift survey of this object to date.  Extensive $N$-body
experiments are performed on simulated superclusters in both
critically-flat and low-density cosmological models to test the
robustness and errors of the mass estimators used on the data.  From
the redshifts and dynamics of 528 galaxies in the supercluster, the
mass of Corona Borealis is estimated to be at least $3\times 10^{16}
\,h^{-1}\,M_\odot$, and the mass-to-light ratio is
$560\,h\,(M/L)_\odot$, yielding a matter density parameter of
$\Omega_m\approx 0.4$.  We also discuss the prospects for mapping the
mass distribution on supercluster scales with gravitational lensing.

\end{abstract}


\section{Introduction}
Among the most remarkable scientific discoveries of the past decades
is the realization that most of the mass in the Universe may reside in
dark matter.  Although ongoing direct searches for dark matter in
particle detectors have not yet revealed positive results, compelling
astrophysical evidence for its existence is now abundant.  Moreover,
various observations now indicate that Nature's ability to hide dark
matter increases with distance scales.  A useful measure of this
``hiding power'' is $M/L$, the ratio of mass to light of an object
expressed in the solar unit $(M/L)_\odot$.  A larger mass-to-light
ratio therefore indicates the presence of more dark matter for the
observed amount of light.  Current measurements give a mass-to-light
ratio of about 5 in the solar neighborhood, 10 in the cores of
elliptical galaxies, 100 for the Local group, and as high as 200-300
for clusters of galaxies (e.g., Carlberg et al. 1996).  Nature's
appetite appears to be more voracious on grander scales, gobbling up
large quantities of matter while exhibiting little sign of digestion.

The motivation for our current study is to obtain measurements of
galaxy and matter distributions on the little-explored supercluster
scales, and to investigate whether the mass-to-light ratio continues
to rise beyond the much-studied cluster scales of a few Mpc.
Superclusters are among the largest structures ever registered on maps
of the nearby galaxies.  Each of these immense clusters of clusters
consists of possibly 10,000 or more galaxies in a region of space tens
of Mpc across.  Our study is focused on the most prominent example of
superclustering in the northern sky -- the Corona Borealis
supercluster, centered at right ascension $15^h20^m$ and declination
$+30^\circ$.  This extraordinary cloud of galaxies was already noted
by Shane and Wirtanen in 1954 and Abell in 1958.  The Corona Borealis
supercluster is now known to consist of seven rich Abell clusters at
$z \approx 0.07$ and numerous galaxies in the intra-cluster regions.
Its core covers a $6^\circ \times 6^\circ$ region of the sky,
corresponding to nearly $20\times 20$ Mpc in physical size.  Whether
the supercluster extends beyond this scale is not yet clear; deeper
and wider surveys of the sky will be required to answer this question.

The previous redshift survey of the Corona Borealis supercluster was
focused on the densest regions surrounding the cores of six individual
Abell clusters A2061, A2065, A2067, A2079, A2089, and A2092 (Postman,
Geller, \& Huchra 1988).  From the dynamics of more than 150 cluster
galaxies, the sum of the masses within the central $1\,h^{-1}$ Mpc of
each of the six clusters was estimated to be $2.4\times 10^{15}
h^{-1}\,M_\odot$.  The mean mass-to-light ratio in the R band for the
same region was about 250.  If the same $M/L$ is assumed to extend to
regions beyond 1 Mpc, the supercluster mass is then $8.2\times 10^{15}
h^{-1}\,M_\odot$.  A higher $M/L$ ratio would further increase the
mass.

\begin{figure}
\epsfxsize=3.2truein 
\epsfbox[0 50 400 620]{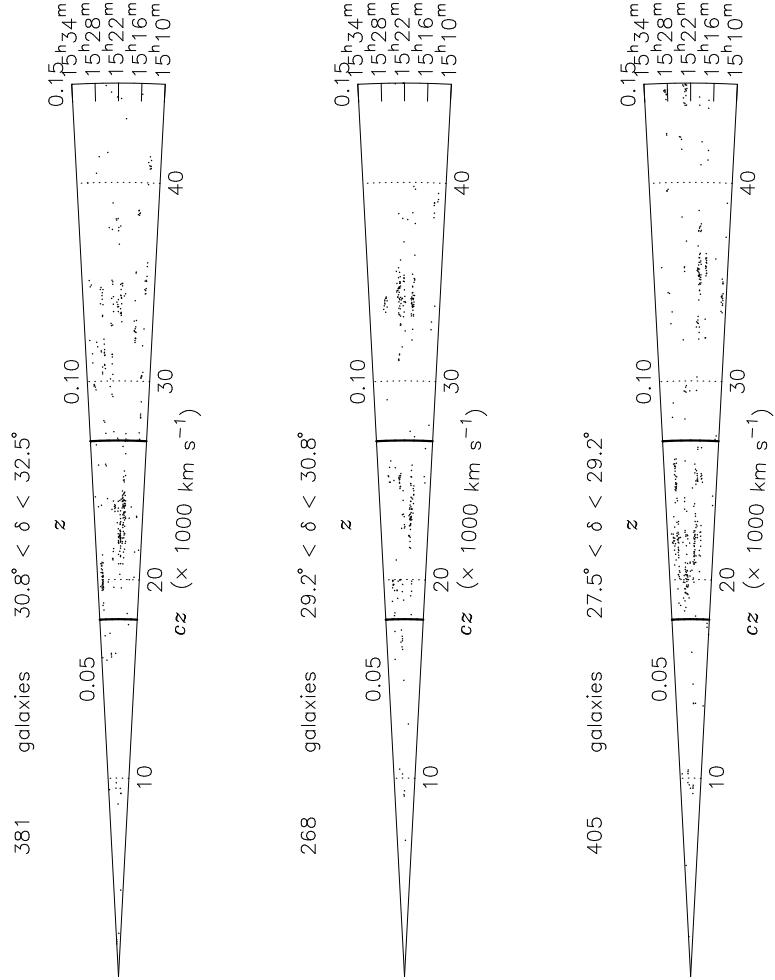}
\caption{Redshift and right-ascension distribution of all galaxies
with redshift $z<0.15$ in our Corona Borealis survey.  The declination
range covered by the survey ($27.5^\circ < \delta < 32.5^\circ$) is
shown in three slices.  The thick lines mark the redshift extent of
the Corona Borealis supercluster, while the dotted lines mark the
background supercluster.}
\end{figure}

A more extensive redshift survey of Corona Borealis has since been
completed with the 176-fiber Norris Spectrograph at the Palomar
200-inch telescope (Small et al. 1997ab; 1998).  This survey
substantially increases the number of galaxy redshifts both in the
Abell clusters and in the intra-cluster regions.  It allows for a more
accurate determination of the structure of the supercluster and a more
reliable estimate of its mass.  A total of 32 fields, each with a
20-arcmin diameter field-of-view, are successfully observed, yielding
redshifts for 1491 extragalactic objects.  With 163 redshifts from the
literature, the entire survey contains 1654 redshifts.  A total of 528
galaxies actually lie in the redshift range of the supercluster, $0.06
< z < 0.09$, and forms the sample for the study reported here on the
supercluster's mass and structure.  Figure~1 shows the redshift and
right-ascension pie diagrams in three declination slices for all
galaxies in our survey with $z<0.15$.  The supercluster is well
delimited along the line of sight by foreground and background voids.
A background supercluster at $z\approx 0.11$ including Abell 2069 is
also visible.  The foreground galaxies at $z\approx 0.03$ are part of
the ``Great Wall'' identified by Geller and Huchra (1989).

\section{Testing Mass Estimators}

The main difficulty in estimating the mass of a supercluster is that
unlike a star or galaxy cluster, a supercluster is so large that the
galaxies and clusters residing in it have not had sufficient time to
interact and randomize their motion.  It is therefore unclear that
traditional estimators such as the virial theorem are valid for a
supercluster.  In order to address this problem, we have chosen to
test the performance of the virial and other mass estimators on
computer-simulated superclusters drawn from large $N$-body simulations
of cosmic structure formation.  The simulated superclusters, which are
in general quite spatially anisotropic, also enable us to assess the
effects of the non-uniform sampling in our observations on our mass
estimates.

To anticipate the possibility that supercluster dynamics has a
systematic dependence on the cosmological density parameter, we have
simulated both critically-flat and low-density models.  The
$\Omega_m=1$ model is the standard cold dark matter (CDM) with a
Hubble constant of $h=0.5$ and a normalization of $\sigma_8=0.7$ for
the rms mass fluctuations in spheres of radius $8\,h^{-1}$ Mpc.  The
low-density model is CDM with $\Omega_m=0.3$, a cosmological constant
$\Omega_{\Lambda}=0.7$, and $h=0.75$, normalized to the 4-year COBE
quadrupole $Q_{\rm rms-PS}=18\,\mu K$ (Gorski et al.\ 1996).  The
corresponding $\sigma_8$ is 1.2.  We performed both large-box
simulations (640 Mpc a side) with random Gaussian initial conditions
and small-box simulations (160 Mpc a side) that are constrained to
produce objects with an overdensity of roughly 5.  A comoving Plummer
force softening length of 160 kpc is used in the particle-particle
particle-mesh (P$^3$M) force calculation for all simulations.

\begin{figure}
\epsfxsize=3.8truein 
\epsfbox[20 130 400 600]{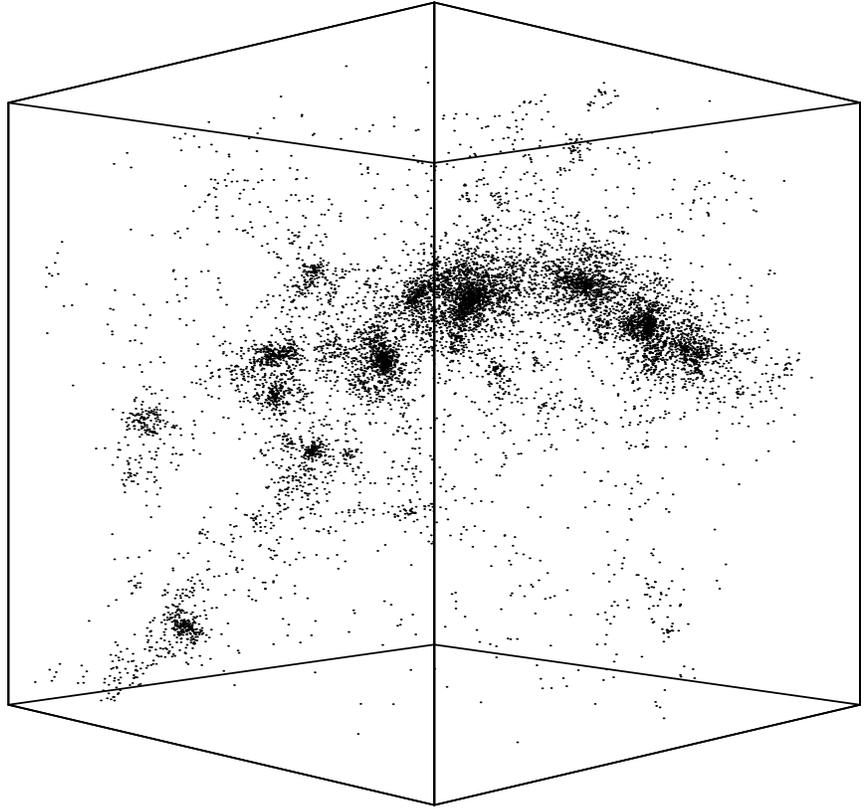}
\caption{An example of the simulated superclusters used to test the
mass estimators applied on the actual data.  This complex structure
consists of at least 7 cluster-type subclumps, each with a mass
exceeding $2\times 10^{14}\,M_\odot$.  Each side of the cubic box is
30 Mpc, and the total mass inside the box is $1.1\times
10^{16}\,M_\odot$.}
\end{figure}

Figure~2 shows the dark matter distribution in one simulated
supercluster.  Each side of the box shown is 30 Mpc, and the total
mass is $1.1\times 10^{16}\,M_\odot$.  This object is extracted from a
large cosmological simulation of the CDM model that traced the motion
of 17 million particles in a (640 Mpc)$^3$ box.  The average
overdensity in this region is $\delta\rho/\rho=4.7$.  We identified 15
other supercluster-like objects that emerged in our cosmological
simulations of both $\Omega_m=1$ and $\Omega_m=0.3$ models.  To
minimize a potential selection bias, only two criteria are applied in
identifying these objects.  First, the candidates must consist of
multiple dark matter halos that have not yet merged into one dominant
smooth halo.  This feature is strongly suggested by the observed
complex substructure (e.g., the Abell clusters) of the Corona
Borealis.  The only other criterion is that the objects have an
average overdensity of $\sim 5$ in a volume of $\sim 30^3$ Mpc$^3$,
corresponding to the least evolved dynamical state of Corona Borealis
(see below).  This is a conservative choice since systems with higher
overdensities are closer to virialization, and the mass estimators
should only work better.

For the virial estimator, the mass is calculated from
\begin{equation}
M_V = {3\pi \over G} \sigma^2 
{1 \overwithdelims \langle\rangle r_p}^{-1},
\end{equation}
where $\sigma$ is the line-of-sight velocity dispersion and $\langle
1/r_p \rangle^{-1}$ is the mean harmonic projected separation.
A simple estimator for $\langle 1/r_p \rangle^{-1}$ is given by
\begin{equation}
{1 \overwithdelims \langle\rangle r_p}^{-1} = {D \over 2}N(N-1)
\biggl(\sum_i \sum_{j < i} {1 \over \theta_{ij}}\biggr)^{-1},
\label{eq:rp}
\end{equation}
where $\theta_{ij}$ is the angular separation of galaxies $i$ and $j$,
$D$ is the radial distance, and $N$ is the total number of galaxies.
This estimator, however, is very sensitive to close pairs and is thus
quite noisy, especially for systems which have not been uniformly
sampled spatially.  We adopt an alternative estimator introduced by
Carlberg et al.\ (1996) that is less sensitive to irregular sampling
and close pairs:
\begin{equation}
{1 \overwithdelims \langle\rangle r_p} = {N(N-1) \over 2}
\sum_i \sum_{j < i} {2 \over \pi (r_i + r_j)} K(k_{ij}),
\end{equation}
where $r_i$ and $r_j$ are the projected radii of objects $i$ and $j$,
$K(k)$ is the complete elliptic integral of the first kind in
Legendre's notation, and $k_{ij}^2 = 4r_ir_j/(r_i+r_j)^2$.  Although
this ``ringwise'' estimator was originally developed for systems such
as galaxy clusters with near circular symmetry on the sky, we find
this estimator to give less biased values of $\langle 1/r_p
\rangle^{-1}$ than the straightforward sum in equation~(\ref{eq:rp}).

We have also tested the projected mass estimator (Bahcall \& Tremaine
1981) given by
\begin{equation}
M_P = {f_{\rm pm} \over G N} \sum_i v_i^2 r_i\,.
\end{equation}
It is designed to give equal weights to particles at all distances (if
$v^2 \propto 1/r$ on the average), but the estimate depends on the
mean eccentricity of the particle orbits parameterized by $f_{\rm
pm}$.  It can be shown that $f_{\rm pm}=32/\pi$ for isotropic orbits
and $64/\pi$ for radial orbits, independent of the mass distribution
(Heisler, Tremaine, \& Bahcall 1985).  We have chosen to use $f_{\rm
pm} = 32/\pi$ since this yields the smallest masses.

Table 1 summarizes our test statistics on the 16 simulated
superclusters in both $\Omega_m=1$ and 0.3 models.  Each supercluster
is projected onto a two-dimensional sky, and the masses are computed
from eqs.~(1) and (4) using all the particles in the supercluster.
Overall, we find that the virial theorem works remarkably well at
recovering the true mass $M_T$ of the simulated superclusters.  An
important result is that the estimators do not perform worse in the
$\Omega_m=0.3$ model as one may have expected for a low-density model.

To test the effects of irregular sampling in our redshift survey, we
have also conducted simulated observations of these model
superclusters with strategies similar to those used during the actual
observations.  Each supercluster is projected onto a two-dimensional
sky and portions of it are viewed through randomly-placed fictitious
observing fields.  Instead of using all particles, a fraction of them
in each field are randomly rejected so that the total number of
particles used in the simulated observations was roughly 500,
comparable to the number of galaxies with measured redshifts in Corona
Borealis.  We find the non-uniform sampling to systematically
underestimate $\langle 1/r_p \rangle^{-1}$ but does not significantly
affect the velocity dispersion.  Overall, the virial estimator
underestimates the true mass by 31\% in the $\Omega_m=1$ model and by
5\% in the $\Omega_m=0.3$ model.  Based on this result and Table 1, we
conclude that the virial and projected mass estimators can be reliably
applied to superclusters even though they are not relaxed systems.
With our irregular sampling, the estimators may underestimate the true
mass by up to 30\%.

\begin{table}
\caption{Test results of mass estimators}
\begin{center}
\begin{tabular}{crr}
Model & $<M_V/M_T>$\tablenotemark{a}  & $<M_P/M_T>$\tablenotemark{b} \\
\tableline
$\Omega_m=1$   & 0.94 $\pm$ 0.24 & 0.76 $\pm$ 0.24 \\
$\Omega_m=0.3$ & 0.99 $\pm$ 0.21 & 1.04 $\pm$ 0.52 \\
\end{tabular}
\end{center}
\tablenotetext{a}{Ratio of virial mass $M_V$ of eq.~(1) to true mass.}
\tablenotetext{b}{Ratio of projected mass $M_P$ of eq.~(4) to true mass.}

\end{table}

An interesting question to ask is why the virial (or similar) mass
estimator works so well for unrelaxed objects with prominent
substructure such as in Figure~2.  An important clue is that our 16
simulated superclusters all turned out to be gravitationally bound
even though this was {\it not} a selection criterion.  How well does
the virial estimator work for a bound system?  For a random collection
of particles, the virial theorem clearly can yield a mass that is
wrong by an arbitrary factor.  For a bound system with $K-|W| <0$ (
where $K$ and $W$ are the kinetic and potential energy), however, it
is easy to show that the virial mass $M_V$ can never exceed twice the
system's true mass $M_T$.  (Recall a virialized system has $K=|W|/2$.)
In fact, $M_V$ approaches $2 M_T$ only for a marginally bound system
with $K\approx |W|$.  For systems with $|W|/2 < K < |W|$, one gets
$M_T < M_V < 2 M_T$; while for systems with $ K < |W|/2$, the virial
estimator underestimates the mass: $M_V < M_T$.  The virial theorem
can therefore recover the true mass of a {\it bound} system to within
a factor of 2 in most cases.  We indeed find this statement to be
true for all 16 simulated superclusters.

Although our $N$-body experiments strongly suggest that superclusters
are generally bound objects, we provide an additional argument here
supporting the possibility that the Corona Borealis supercluster is
gravitationally bound.
The spherical tophat model for gravitational collapse indicates that
an outer shell of a density perturbation is bound if the mean
overdensity within the shell is greater than $(\Omega_m^{-1}-1)/(1+z)$
(e.g., Peebles 1980).  Taking 0.3 to be a rough lower limit on
$\Omega_m$, a supercluster at $z=0.07$ must be overdense by more than
a factor of 2.1 for it to be bound.  In comparison, the observed
overdensity in galaxy counts in Corona Borealis relative to the field
is $\delta\approx 7f$, where the dimensionless parameter $f$, defined
by $f\equiv c\,\delta z/H_0\,\delta r$, measures the relative
elongation in redshift space $\delta z$ and real space $\delta r$
along the line of sight.  The minimal value of $f$ is unity,
corresponding to zero peculiar velocities, whereas a larger $f$
indicates a stronger ``finger-of-god'' effect due to galaxy peculiar
motion.  Although we cannot completely rule out the $f=1$ case, it is
very difficult to arrange an object that is as immense and prominent
as the Corona Borealis while at the same time is freely expanding with
the Hubble flow without exhibiting any peculiar velocities.  The
overdensity in galaxy counts in the Corona Borealis is therefore at
least 7, but more likely to be significantly above this value.  The
Corona Borealis supercluster is therefore bound unless galaxies are
significantly biased relative to mass by more than a factor of 3.5.

\section{Mass and $M/L$ of Corona Borealis}

Given the robustness of the mass estimators found in our numerical
tests described in the previous section, we now proceed to apply
eqs.~(1)-(4) to the actual data from the Corona Borealis survey.  Using
the techniques of Beers, Flynn, \& Gebhardt (1990), we estimate the
centroid velocity of the Corona Borealis supercluster to be $c\bar z =
22420^{+149}_{-138}$ km s$^{-1}$ and the dispersion to be $\sigma =
1929^{+81}_{-67}$ km s$^{-1}$ in the cluster rest frame.  Using the
ringwise estimator in eq.~(3), we find $\langle 1/r_p \rangle^{-1} =
4.6\,h^{-1}$ Mpc.  The virial mass estimator (eq.~(1)) then gives a
mass of $3.8 \times 10^{16}\,h^{-1}$ M$_\odot$ for the Corona Borealis
supercluster.  The projected mass estimator in eq.~(4) yields a
similar value, $4.2 \times 10^{16}\,h^{-1}$ M$_\odot$.  Based on the
tests described in the previous section, we conclude that a secure
lower bound to the mass of the Corona Borealis supercluster is $3
\times 10^{16}\,h^{-1}$ M$_\odot$.

The supercluster luminosity function for $M(B_{AB}) \le -16.3 + 5 \log
h$ mag in the AB-normalized B band is presented in Small et
al. (1997b).  Integration of the luminosity function yields a mean
luminosity density of $\rho_L(B_{AB}) = 1.9 \times 10^9\,h\,L_\odot$
Mpc$^{-3}$ in the supercluster.  Taking the solid angle of the survey
to be 0.0076 sr (= 25 deg$^2$) and the redshift limits of the
supercluster to be at $z = 0.06$ and 0.09, the volume of the region
surveyed is $2.8 \times 10^4\,h^{-3}$ Mpc$^3$.  The $M/L$ ratio of the
supercluster in the $B_{AB}$ band is thus $560\,h\,(M/L)_\odot$.  For a
local $B_{AB}$-band luminosity density of $(1.8 \pm 0.2) \times
10^8\,h\,L_\odot$ Mpc$^{-3}$ for $M(B_{AB}) < -16 + 5 \log h$ mag
(Small et al. 1998 and references therein), the corresponding critical
$M/L$ ratio to close the universe is $1550 \pm 170\,h\,(M/L)_\odot$.
We therefore obtain $\Omega_m = 0.36$ for the density parameter on
supercluster scales of $\sim 20\,h^{-1}$ Mpc.

\section{Future Prospects}
Our determination of $M/L$ and $\Omega_m$ on supercluster scales can
be strengthened by measurements of other superclusters with techniques
similar to those described above, or by independent measurements with
alternative mass-estimating methods.  The former will be achieved when
the ongoing large-area redshift surveys (such as the 2dF and Sloan
surveys) generate sufficient data.  For the latter, a very powerful
technique is gravitational lensing, which probes the mass profile of a
foreground object by the gravitational distortion it induces on the
images of background objects.  A galaxy cluster, for example, induces
a correlated elliptical distortion in the images of background
galaxies.  Measurements of such statistical signals over many galaxies
have been used successfully to map the surface mass densities of
individual clusters (e.g., Tyson et al. 1990; Kaiser \& Squires 1993;
Mellier et al. 1996 and references therein).  Extending the
gravitational lensing technique to superclusters will be a promising
alternative probe of the dark mass distribution on such large length
scales.

\begin{figure}
\epsfxsize=3.5truein 
\epsfbox[72 420 400 680]{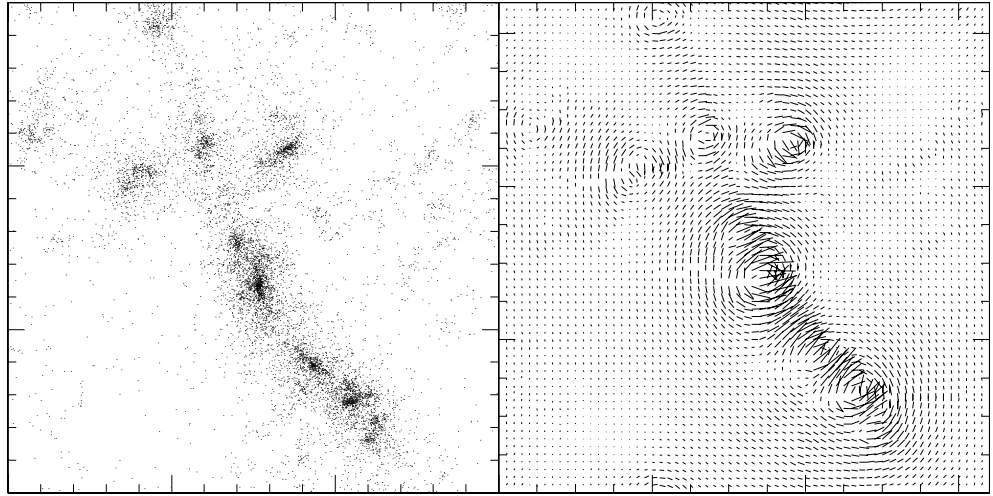}
\caption{Left: A different projection of the simulated supercluster
shown in Figure~2.  Each side of the box is 30 Mpc, and the total mass
in the box is $1.1\times 10^{16}\,M_\odot$.  Right: The local
magnitude and orientation of the shear field induced by the
supercluster.}
\end{figure}

Figure~3 is an illustration of the lensing signal expected in a
superclustering region.  The left panel shows the particle
distribution in a different projection of the same simulated
supercluster in Figure~2.  Each bar in the right panel shows the local
strength and direction of the theoretical elliptical distortion in the
shapes of the background galaxies due to lensing by the supercluster.
More explicitly, if the two-dimensional gravitational potential is
denoted by $\psi\,$, and $\gamma_1\equiv
(\partial_x^2-\partial_y^2)\psi/2$ and $\gamma_2\equiv
\partial_x\partial_y\psi$, then the magnitude of each bar is
proportional to the shear $\gamma\equiv (\gamma_1^2+\gamma_2^2)^{1/2}$
that measures the anisotropic stretching of the images.  The angle
$\beta$ of each bar relative to the horizontal axis obeys
$\tan(2\beta)=\gamma_2/\gamma_1$.  Comparison of the two figures shows
that the amplitude of the distortion on the right is clearly
correlated with the projected mass density on the left, and the
preferential tangential alignment of the shear orientation is evident.
A promising feature is that the shear amplitudes do not drop
appreciably along the ridges connecting individual cluster-like
clumps.  Measurements of the distortion in the images of background
galaxies can therefore in principle be used to construct a surface
mass density map in a supercluster.  This offers a powerful
alternative way to probe the distribution of dark matter on scales of
tens of Mpc.

\acknowledgments 
We are grateful to the Kenneth T. and Eileen L. Norris Foundation for
their generous grant and Donald Hamilton for construction of the Norris
Spectrograph.  We thank Jim Frederic and Paul Bode for aid with the
$N$-body simulations, and Roger Blandford and David Buote for helpful
comments.  Supercomputing time was provided by the National Scalable
Cluster Project at the University of Pennsylvania, the National Center
for Supercomputing Applications, and the Cornell National
Supercomputer Facility.  This work has been supported by an NSF
Graduate Fellowship (TAS), a Caltech PMA Division Fellowship (CPM),
and NSF grant AST92-213165 (WLWS).

\end{document}